\documentclass[onecolumn,showpacs,preprintnumbers,amsmath,amssymb]{revtex4}

\usepackage{amsbsy}
\usepackage{graphicx}

\begin{document}

\title{Domain Walls in a Tetragonal  Chiral $p$-Wave Superconductor}

\author{N.~A.~Logoboy and E.~B.~Sonin}

\affiliation{Racah Institute of Physics, Hebrew University of
Jerusalem, Jerusalem 91904, Israel}

\date{\today}

\begin{abstract}

Domain walls in a tetragonal chiral $p$-wave superconductors
 with broken time reversal symmetry  are analyzed in the framework of the Ginsburg-Landau theory.
The energy and the jump of the magnetic induction on the wall were determined for different types of walls as functions of the parameters of the Ginzburg-Landau theory and orientation of the domain wall with respect to the crystallographic axes.  We discuss implications of the analysis for  $\textrm{Sr}_{2}\textrm{RuO}_{4}$, where no stray magnetic fields from domain walls were detected experimentally.
\end{abstract}


\maketitle

\section{\label{sec:Introduction}Introduction}

There were a number experimental evidences of the broken time-reversal symmetry (TRS) in the unconventional superconductor $\textrm{Sr}_{2}\textrm{RuO}_{4}$ \cite{Luke,ishida,Xia}.
It was suggested that  this phenomenon was connected with  the $p$-wave Cooper  pairing  with the wave function in the momentum space proportional to $p_x + ip_y$ (chiral p-wave
superconductivity) \cite{Sigrist2,Mackenzie}.  In chiral  $p$-wave superconductors the spontaneous magnetic flux must be present near domain walls or sample boundaries. However, Kirtley {\em et al.} \cite{Kirtley} have not detected any  stray magnetic fields, which the spontaneous magnetic flux should produce  above the sample surface. This put in question the scenario of the chiral $p$-wave pairing and stimulated theoretical investigations of the problem. In particular, the relation between the stray fields and the magnetic flux, which appears near the domain wall (DW) in the bulk, was derived \cite{Bluhm,Logoboy}. Another  challenge for the theory was to find the spontaneous magnetic flux itself. It is determined by the product of the London penetration depth and the jump of the magnetic induction on the domain wall. The latter is on the order of the first critical magnetic field $H_{c1}$  and was already analyzed in the past for particular types of the DWs in chiral superconductors \cite{Volovik,Sigrist1,Matsumoto}.

There are two possible explanations why the experiment could not detect the stray fields above the sample surface. The first one  is that there is a domain structure with a period so small that stray fields decay very fast in  space at a distance small compared to the distance of the experimental probe from the sample surface. There were some experimental evidences of domain structure in $\textrm{Sr}_{2}\textrm{RuO}_{4}$ \cite{Tamegai,Kidwingira}, though Xia {\em et al.} \cite{Xia} did not reveal any domain  structure at studying the Kerr effect. The theory predicts that in superconductors with broken TRS  usual ferromagnetic domains, which depend on the size and the shape of the sample (extrinsic domains), cannot appear \cite{Sonin}, but another type of domains, which decrease the bulk magnetostatic energy at the expense of destroying the Meissner state, becomes possible \cite{Krey,Faure}.
The size of these domains is roughly of the order of the London penetration depth $\lambda$ and does not depend on either shape or size of the sample (intrinsic domains, see discussion in Ref.~\onlinecite{Sonin2}). The analysis of  intrinsic domain structure in superconductors with broken TRS was recently extended on the case of finite external magnetic fields, and the equilibrium magnetezation curves in the state with intrinsic domains (cryptoferromagnetic state) were found  theoretically \cite{LS}. The intrinsic domain structure may lead to serious suppression of stray fields above the sample, though definite quantitative conclusions are not yet possible because of the absence of essential information on $\textrm{Sr}_{2}\textrm{RuO}_{4}$ (magnetic crystal anisotropy, e.g.). Therefore, the other possible explanation of the negative result of Kirtley {\em et al.} \cite{Kirtley} must be also considered: a very small  spontaneous magnetic flux  penetrating along DWs in $\textrm{Sr}_{2}\textrm{RuO}_{4}$.  This problem is investigated in the present work.

The magnetic flux near DWs originates from the intrinsic orbital moment of Cooper pairs (orbital ferromagnetism). As was demonstrated in Ref.~\onlinecite{Braude1}, in the case of orbital ferromagnetism one should not use the Landau-Lifshitz theory \cite{LL}, since in this case one cannot define local spontaneous magnetization. Therefore one should rely on magnetization currents generated by the orbital moment of Cooper pairs, which cannot be reduced to the curl of any vector moment. The total magnetization current along the DW leads to the jump of the magnetic induction on the DW, which eventually determines the magnetic flux connected with the DW. In the present work we extend  previous theoretical investigations of the magnetic flux around DWs in tetragonal chiral $p$-wave superconductors analyzing conditions for appearance of different types of DW structure depending on the strength of crystal in-plane anisotropy and DW orientation with respect to crystallographic axes.
The analysis was done using the Ginzburg-Landay (GL) theory, and the magnetic induction near DWs depends on the parameters of the GL theory. For the parameters obtained in the weak-coupling limit the magnetic induction near DWs is scaled by the first critical magnetic field $H_{c1}$ differing from the latter by a numerical factor. But for other values of the GL parameters the magnetic induction near DWs can be much smaller and even become negative with respect to the intrinsic magnetic moment of Cooper pairs. This stresses again that the orbital moment of the Cooper pair in the chiral $p$-wave state does not lead to any definite local magnetization.

The practical outcome of the presented analysis is that one can explain very weak stray fields from DWs if the GL parameters differ from those calculated in the weak-coupling limit. The final conclusion on the reason for the absence of detectable stray fields around $\textrm{Sr}_{2}\textrm{RuO}_{4}$ samples is possible  only after  these parameters are found from the experiment or from the strong-coupling theory.

\section{\label{sec:Ginsburg-Landau}Model}

The unconventional superconductor  Sr$_{2}$RuO$_{4}$ belongs to the  tetragonal crystallographic symmetry group
$D_{4h}$. Considering the $p$-wave state for this material they usually believe that strong crystal anisotropy keeps both spin and orbital momentum of the Cooper pair parallel to the $c$ axis. Then the $p$-wave state corresponds
to the two-dimensional representation $\Gamma^{-}_{5}=\{k_{x}\mathbf{\hat{z}},k_{y}\mathbf{\hat{z}}\}$, and the order parameter  $(\eta_x k_x+\eta_y k_y)\mathbf {\hat{z}}$  is
fully described by a complex two-component  vector ${\mathbf \eta}=(\eta_{x},\eta_{y})$ \cite{Mackenzie,Sigrist2}. Then the GL free energy density is
\begin{widetext}
\begin{eqnarray}\label{eq:GL1}
\mathcal {F}=P_{1}\vert\mathbf{\eta}\vert
^{2}+\beta_{1}\vert\mathbf{\eta}\vert^{4}+\beta_{2}(\eta^{\ast}_{x}\eta_{y}-\eta_{x}\eta^{\ast}_{y})^{2}
+\beta_{3}\vert\eta_{x}\vert^{2}\vert\eta_{y}\vert^{2}+K_{1}\vert\mathbf{D}_{x,y}\cdot\mathbf{\eta}\vert^{2}+K_{2}(\vert
D_{x}\eta_{y}\vert^{2}+\vert D_{y}\eta_{x}\vert^{2}) \qquad \qquad \nonumber \\
+K_{3}[(D_{x}\eta_{x})^{\ast}(D_{y}\eta_{y})+(D_{x}\eta_{x})(D_{y}\eta_{y})^{\ast}]+K_{4}[(D_{x}\eta_{y})^{\ast}(D_{y}\eta_{x})+(D_{x}\eta_{y})(D_{y}\eta_{x})^{\ast}]+
K_{5}(\vert D_{z}\eta_{x}\vert^{2}+\vert D_{z}\eta_{y}\vert^{2}).
\quad
\end{eqnarray}
\end{widetext}
Here  $D_{i}=\partial_{i}-i(2e/c\hbar)A_{i}$ is a covariant
derivative and $\mathbf{B}=\mathbf{\nabla}\times\mathbf{A}$ is the magnetic induction.
In the case of full axial symmetry in the plane $\beta_3=0$ and $K_1=K_2+K_3+K_4$, In the BCS theory (weak-coupling limit) $K_1/3=K_2=K_3=K_4$ \cite{WC}.

The complex components of the  two-component  order
parameter $\mathbf{\eta}$ can be represented in a form
\begin{eqnarray}
  \eta_{x}= \eta_{0}~\cos \Theta~ \exp {(i \phi)},\nonumber   \\
  \eta_{y}= \eta_{0}~\sin \Theta~ \exp {[i (\phi+\zeta)]},
  \label{eq:General Expression 2C OP}
\end{eqnarray}
where angles (phases) $\Theta,~\phi $, and $\zeta$ are coordinate
dependent functions in general, and $\eta_{0}=|\mathbf{\eta}|$. In the
focus of the present study  is the chiral state with the
two-fold degenerate ground state $\mathbf {\eta} =
2^{-1/2}\eta_{0}\exp{(i\phi)} (1,\pm i)$, which corresponds to $\Theta=\pi/4$ and $\zeta=\pm \pi/2$ and  is realized if
$\beta_{2}>0$ and $\beta_{1}> \beta_{2}-\beta_{3}/4>0$. The
order-parameter amplitude,
\begin{equation} \label{eq:Order Parameter}
  \eta^{2}_{0}=\frac{2\mid P_{1}\mid}{4(\beta_{1}-\beta_{2})+\beta_{3}},
\end{equation}
is determined by minimization of the GL free energy. The chiral state has a nonzero orbital moment ${\mathbf l}=i {\mathbf \eta}\times
{\mathbf \eta}^{\ast}=l_{z} \hat{z}$ with the only component
 \begin{equation} \label{eq:Orbital Moment}
l_z=i(\eta_{x}\eta^{\ast}_{y}-\eta_{y}\eta^{\ast}_{x})= \eta^{2}_{0}\sin 2\Theta \sin \zeta .
\end{equation}
Other components of the orbital moment are suppressed by strong crystal anisotropy. Though they are necessary for orbital magnons \cite{Braude1}, they can be ignored in our analysis of DWs.

Substituting the general expression for the two-component order
parameter, Eq.~(\ref{eq:General Expression 2C OP}),  into the free energy density
$\mathcal {F}$ [Eq.~({\ref{eq:GL1})], one can derive the
expression for the superconducting  current:
\begin{equation} \label{eq:Current Density}
\mathbf{j}=\frac{e}{m}\partial_{\mathbf{v}}\mathcal
{F}=\mathbf{j}^{(tr)}+\mathbf{j}^{(m)}, \qquad
\end{equation}
which consists of two contributions.  The
transport current $\mathbf{j}^{(tr)}=e\hat { n} {\mathbf v}$
is determined by the gauge-invariant superfluid velocity
\begin{equation} \label{eq:Velocity}
\mathbf{v}=\frac{\hbar}{2 m}\mathbf{\nabla}\phi-\frac{e}{mc}\mathbf{A},
\quad
\end{equation}
where  $\hat n$ is the superfluid electron-density matrix  with the components
\begin{eqnarray}
n_{xx}={8m\over \hbar^2}\eta_{0}^2(K_{1}\cos^{2}\Theta +K_{2}\sin^{2}\Theta), \nonumber \\
n_{yy}={8m\over \hbar^2}\eta_{0}^2(K_{1}\sin^{2}\Theta+K_{2}\cos^{2}\Theta), \nonumber \\
n_{xy}=n_{yx}={4m\over \hbar^2}\eta_{0}^2(K_{3}+K_{4})\sin 2\Theta
\cos\zeta, \nonumber \\
n_{zz}={8m\over \hbar^2}\eta_{0}^2 K_{5}.
 \label{eq:Conductivity Tenzor}
\end{eqnarray}
The second contribution to the current is the magnetization current with the components
\begin{eqnarray}\label{eq:Magnetization Current 1}
j^{(m)}_{x}={2e\over \hbar}\eta_{0}^2[2 K_{2}\sin^{2}\Theta~\partial_{x}\zeta+2(K_{3}\cos^{2}\Theta+
K_{4}\sin^{2}\Theta)\sin\zeta~\partial_{y}\Theta+
K_{3}\sin 2\Theta\cos\zeta~\partial_{y}\zeta], \nonumber\\
j^{(m)}_{y}={2e\over \hbar}\eta_{0}^2[2 K_{1}\sin^{2}\Theta~\partial_{y}\zeta+2(K_{3}\sin^{2}\Theta+
K_{4}\cos^{2}\Theta)\sin\zeta~\partial_{x}\Theta+
K_{4}\sin 2\Theta\cos\zeta~\partial_{x}\zeta], \nonumber\\
j^{(m)}_{z}={4e\over \hbar}\eta_{0}^2K_{5}\sin^{2}\Theta~\partial_{z}\zeta.
\end{eqnarray}
In the following we shall neglect the terms $\propto K_5$ considering only DWs parallel to the axis $z$ when parameters do not vary along this axis. We also assume that the London penetration depth $\lambda$ is much larger than the thickness of the DW, which is on the order of the coherence length $\xi$. This allows to ignore Meissner currents studying currents inside the DW.

Inside domains $\Theta=\pi/4$, and $n_{xx}=n_{yy}=(4 m/\hbar^2)\eta_{0}^2(K_{1}+K_{2})$. Then the London penetration depth  and the first critical magnetic field are
\begin{equation}
\lambda=    \sqrt{\hbar^2 c^2\over 16\pi e^2\eta_0^2 (K_1+K_2)},~~
H_{c1} ={4\pi e\over \hbar c}  \eta_0^2(K_1+K_2)\ln {\lambda \over \xi}.
 \label{Hc1}
\end{equation}

The magnetization current cannot be presented as $\mathbf{j}^{(m)} =
(1/c) {\mathbf \nabla} \times {\mathbf  M}$, which is the main
assumption of the Landau-Lifshitz theory. Therefore for chiral
superconductors the Landau-Lifshitz theory is not valid, and the
local magnetization ${\mathbf  M}$ cannot be defined \cite{Braude1}.
Whereas in the Landau-Lifshitz theory the jump of  the tangential
component of the magnetic  induction on the DW is given by the
universal relation $\delta B=8\pi M$ independently of the DW
structure, in chiral superconductors  $\delta B $ does depend on the
DW type and on  the DW orientation relative to crystallographic
axes. This will be demonstrated in the next section.

\section{Domain walls}
\label{Results and Discussion}

The two degenerate  ground states correspond to two possible
directions of the orbital moment $l_z=\pm 1$ parallel or
anti-parallel to the $c$ axis (the axis $z$). This   leads to the existence of
domains with $l_z=\pm 1$ separated by domain walls.
Two types of DWs  are known. The DW of the type I is characterized by a gradual change of
the phase $\zeta=\zeta(\mathbf{r})$ at constant $\Theta=\pi/$4, i.e., the absolute values of the components $|\eta_x|=|\eta_y|$ remain constant inside the DW  \cite{Volovik}.
 We address this type of the DW as
$\zeta-$wall. In the DW of the type II the relative phase $\zeta=-
\pi/2$ remains fixed, and the transition from $l_z=-1$ to $l_z=+1$
is realized via variation of the angle $\Theta$ so that in the
center of the DW one of the components $\eta_x$ or $\eta_y$
vanishes. This type of DW was considered by Sigrist {\em et al.}
\cite{Sigrist1} assuming that the component of $\mathbf \eta$, which
does not vanish, remains constant. In the present work we assume
that inside the DW the order parameter $\mathbf \eta$ rotates in
space at fixed order parameter modulus $|\mathbf{\eta}|=\eta_0$.
This assumption is justified if $\beta_1 \gg \beta_2,\beta_3$.
Further we address this type of the DW as $\Theta-$wall. However,
the difference between the two types of DW depends on a choice of
the coordinate system: rotating the in-plane coordinate frame $xy$,
in which the components of $\mathbf \eta$ are defined, through the
angle $\pi/4$ (45$^\circ$), the $\zeta$-wall becomes a $\Theta$-wall
and vice versa.

\subsection{\label{GCofP:GC}Variation of relative phase: $\zeta$-wall}

Let us consider a DW,  which is characterized by continuous variation of the relative
phase $\zeta=\zeta(\mathbf{r})$ at fixed angle  $\Theta=\pi/$4. The transformation of the $l_z=-1$ domain to   the $l_z=+1$ domain
is possible via counterclockwise  rotation of $\zeta$ from $-\pi/2$ to $\pi/2$ or clockwise  rotation of $\zeta$ from $-\pi/2$ to $-3\pi/2$.
 The structure of this
DW, its surface energy, and the jump of magnetic
induction depend on its orientation with respect  to chosen crystallographic axes $(xy)$.
Assuming the
arbitrary orientation of the DW plane in the $xy$-plane we
transform the gradient terms in Eq.~(\ref{eq:GL1}) to the new
coordinate system $(x'y')$ connected to the DW: the axis $x'$ is normal and the axis $y'$ is parallel to the DW:
\begin{eqnarray}
\partial_x=\cos\psi \partial_{x'}-\sin\psi\partial_{y'},
\nonumber \\
\partial_y=\sin\psi \partial_{x'}+\cos\psi\partial_{y'}.
      \end{eqnarray}
The transformation is a counterclockwise rotation of the original
coordinate axes $xy$ around $z$-axis through an angle $\psi$ about the origin. Note that only the gradients were rotated, the order parameter ${\mathbf \eta}$ being untouched and characterized by the components $\eta_{x,y}$ in the original coordinate system $xy$.
After the transformation the free energy density with respect to the energy of the uniform chiral phase is
\begin{eqnarray}
\Delta \mathcal {F}= {\beta_2 \eta_0^4\over 2}(1+\cos 2 \zeta)+{K_1\eta_{0}^2\over 2}\{  (\cos\psi D_{x '} \phi-\sin\psi  D_{y'} \phi) ^2+[\sin\psi (D_{x '} \phi+\partial_{x '}\zeta)+\cos\psi(D_{y '} \phi+\partial_{y '} \zeta)]  ^2\}
\nonumber \\
 +{K_2\eta_{0}^2\over 2}\{ (\sin\psi D_{x '} \phi+\cos\psi D_{y'} \phi) ^2+ [\cos\psi (D_{x '} \phi+\partial_{x '} \zeta)-\sin\psi\partial_{y'}( \phi+\zeta)] ^2\}
\nonumber \\
+K_3 \eta_{0}^2\cos\zeta (\cos\psi D_{x '} \phi-\sin\psi D_{y'} \phi) [\sin\psi (D_{x '} \phi+\partial_{x '} \zeta)+\cos\psi(D_{y '} \phi+\partial_{y '} \zeta)]
\nonumber \\
 +K_4\eta_{0}^2\cos\zeta (\sin\psi D_{x '} \phi+\cos\psi\partial_{y'} \phi) [\cos\psi(D_{x '} \phi+\partial_{x '} \zeta)-\sin\psi(D_{y '} \phi+\partial_{y '} \zeta).
       \end{eqnarray}
The relevant
components of the superfluid electron-density matrix are
\begin{eqnarray}
 n_{x'x'}={4 m\over \hbar^2}\eta_{0}^2[K_{1}+K_{2}+(K_{3}+K_{4})\sin 2\psi \cos\zeta], \nonumber \\
 n_{x'y'}=n_{y'x'}={4 m\over \hbar^2}\eta_{0}^2(K_{3}+K_{4})\cos 2\psi \cos \zeta.
 \label{eq:Conductivity Tenzor: Type I}
\end{eqnarray}
The magnetization currents are  defined by
\begin{eqnarray}
j^{(m)}_{x'}={e\over  \hbar}\eta_{0}^2[K_{1}+K_{2}-(K_{1}-K_{2})\cos 2\psi+(K_{3}+K_{4})\sin 2\psi \cos\zeta]\partial_{x'}\zeta, \nonumber  \\
j^{(m)}_{y'}={e\over  \hbar}\eta_{0}^2\{[-K_{3}+K_{4}+(K_{3}+K_{4})\cos
2\psi]\cos\zeta+(K_{1}-K_{2})\sin 2\psi\}\partial_{x'}\zeta.
 \label{eq:Magnetization Current Type I}
\end{eqnarray}
The current normal to the DW plane must vanish:
$j_{x'}=e n_{x'x'}v_{x'}+ j^{(m)}_{x'}=$0. This gives the following
expression for $x'$-component of the velocity:
\begin{equation} \label{eq:Velocity Type I}
v_{x'}={\hbar\over 2m} D_{x '} \phi =-\frac{\hbar}{4m}\left[1-\frac{(K_{1}-K_{2})\cos
2\psi}{K_{1}+K_{2}+(K_{3}+K_{4})\sin 2\psi \cos
\zeta}\right]~\partial_{x'}\zeta.
\end{equation}
After its exclusion the total current $j_{y'}$ parallel to the DW is
\begin{equation}
 j_{y'}={e\over \hbar}\eta_{0}^2\left[-(K_{3}-K_{4})\cos \zeta+(K_{1}-K_{2})\frac{(K_{1}+K_{2})\sin 2\psi+(K_{3}+K_{4})\cos \zeta}{K_{1}+K_{2}+(K_{3}+K_{4})\sin 2\psi\cos \zeta}\right]~\partial_{x'}\zeta.
 \label{eq:Total Current Type I}
\end{equation}
Using Eq.~(\ref{eq:Total Current Type I}) and the Maxwell equation one can calculate the jump
of the magnetic induction on the $\zeta$-wall:
\begin{equation}
 \delta B_{\zeta}(\psi) =-(4\pi/c)\int dx' j_{y'}=\frac{8\pi  e}{\hbar c}\eta_{0}^2\left\{ (K_{3}-K_{4}) \mp \frac{K_{1}-K_{2}}{\sin 2\psi} \left[\frac{\pi}{2}-\frac{2\cos^{2} 2\psi }{(1-q^{2})^{1/2}}\arctan\sqrt{\frac{1\mp q}{1\pm q}}\right]\right\},
 \label{eq:Jump of magnetic induction Type I}
\end{equation}
where
\begin{equation}
  q=\frac{K_{3}+K_{4}}{K_{1}+K_{2}}\sin 2\psi.
                                        \end{equation}
The upper and the lower signs correspond to $\zeta$ rotations from $-\pi/2$ to $\pi/2$ or to $-3\pi/2$ respectively.

After the  exclusion of the transport velocity  the free energy density  becomes
\begin{eqnarray}
\Delta \mathcal {F}= {\beta_2 \eta_0^4\over 2}(1-\cos 2 \gamma)+(K_1+K_2)\eta_{0}^2 f (\gamma){\partial_{x'}\gamma^2 \over 8},
  \label{en-xi}       \end{eqnarray}
where
\begin{equation}
 f(\gamma)=1\mp q \sin \gamma-\left(\frac{K_{1}-K_{2}}{K_{1}+K_{2}}\right )^{2}{\cos^{2}2\psi \over 1\pm q \sin \gamma}.
 \label{eq:Notation Structure Type I}
\end{equation}
Here we introduced the angle $\gamma $ equal to $\gamma =\zeta +\pi/2$ or $\gamma =-\zeta -\pi/2$
for $\zeta$--rotations from $-\pi/2$ to $\pi/2$ or to $-3\pi/2$ respectively. So across the DW $\gamma$ varies from 0 to $\pi$.

The structure of the $\zeta$-wall is determined by the Euler-Lagrange equation obtained by variation of the free energy
with the  density (\ref{en-xi}) with
respect to $\gamma$:
\begin{equation}
f(\gamma)\partial_{x'}^2 \gamma +{1\over 2}f(\gamma )'(\partial_{x'} \gamma)^2 -{\sin 2 \gamma \over \Delta^2} =0,
\label{EL} \end{equation}
where the scale
\begin{equation}
 \Delta=\frac{1}{2\eta_{0}}\left(\frac{K_{1}+K_{2}}{\beta_{2}}\right)^{1/2}
 \label{eq:DW Prameter Type I}
\end{equation}
determines the thickness of the DW.
The first integration of Eq. (\ref{EL}) yields
\begin{equation}
(\partial_{x'} \gamma)^2 = \frac{1-\cos 2 \gamma }{\Delta^2f(\gamma)}.
     \end{equation}
After this one can find the surface energy of the DW:
\begin{equation}
 \sigma_{\zeta}(\psi)= \eta^3_{0}\sqrt{\beta_{2}(K_{1}+K_{2})\over 2}\int^\pi_0d\gamma  \sin \gamma \sqrt{f(\gamma)}.
 \label{eq:DW Energy Type I}
\end{equation}

Figure~\ref{Fig_1} shows the plots of the magnetic induction jump $\delta B_\zeta$ and the surface energy as functions  of the angle $\psi$ between  the DW  and one  of the crystallographic axes  for the weak-coupling case when $K_{2}=K_{3}=K_{4}=K_{1}/3$.  The plots for the clockwise
rotation ($\zeta$ varies within DW from $-\pi/2$ to $-3\pi/2$, curves 2) are obtained from
 the counterclockwise
rotation ($\zeta$ varies within DW from $-\pi/2$ to $\pi/2$, curves 1) by reflection $\psi \to -\psi$.
The angular dependences have two extrema at $\psi =\pm \pi/4$.  The spatial distribution  of the magnetization currents along the DW  for different orientations of
the DW is shown in Fig.~\ref{Fig_2}.

\begin{figure}[t]
  \includegraphics[width=0.4\textwidth]{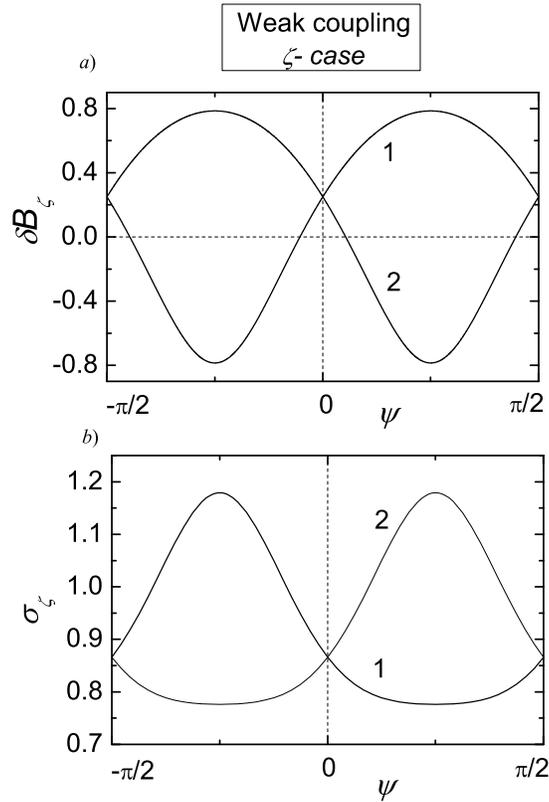}
\caption{\ Shown are the angular dependence (a) of the jump of
reduced magnetic induction $\delta B_{\zeta} \times 3 \hbar c/32 \pi
e \eta^{2}_{0}K_{1}$ and (b) the reduced surface energy
$\sigma_{\zeta} \times 3/4\eta^{3}_{0}(2\beta_{2}K_{1})^{1/2}$ of
the DW for two configurations, when $\zeta$ increases from $-\pi/$2
to $\pi/$2 (curve 1), or decreases from $-\pi/$2 to $-3\pi/$2 (curve
2) with increase of $x'$.} \label{Fig_1}
\end{figure}

\begin{figure}[t]
  \includegraphics[width=0.4\textwidth]{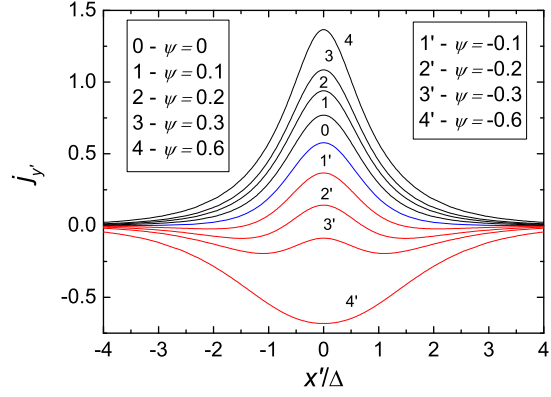}
\caption{\ (color online) The coordinate dependence of the reduced
superconducting current $j_{y^{'}} \times 3^{1/2}\hbar /2^{3/2}e
\eta^{3}_{0}(K_{1}\beta_{2})^{1/2}$ at different orientations of the
DW ($|\zeta|\le \pi/2$).} \label{Fig_2}
\end{figure}

Let us consider some important particular cases. The case $\psi=$0 corresponds to the DW parallel to one from chosen crystallographic axes. The surface energy
and the jump of magnetic induction in this case are
\begin{eqnarray} \label{eq:SSE and Jump I}
\sigma_\zeta(0)=2\eta^3_{0}\sqrt{2\beta_{2}K_1 K_2\over K_1+K_2}, \qquad \quad \nonumber \\
\delta B_\zeta(0)=\frac{16\pi  e}{\hbar c}\eta_{0}^2\frac{K_{2}K_{3}-K_{1}K_{4}}{K_{1}+K_{2}}.
 \end{eqnarray}
This agrees with the results of Volovik and Gor'kov \cite{Volovik}, who assumed that $K_3=K_4$.
In the case $\psi=0$ there is degeneracy between DWs corresponding to two senses of rotation of the phase $\zeta$: the surface energies and the field jumps for two cases coincide.

Another particular case is the angle $\psi=\pi/4$, where the plots in Fig.~\ref{Fig_1} have extrema.  In this case according to Eq.~(\ref{eq:Jump of magnetic
induction Type I}) the jump of magnetic induction is
\begin{equation}
\delta B_{\zeta}\left(\pi\over 4\right)=\frac{8\pi e}{c}\eta^{2}_{0}\left[K_3-K_4 \mp \frac{\pi}{2}(K_{1}-K_{2})\right],
 \label{eq:Maximum Jump}
\end{equation}
and is different for both configurations. The DW surface energy can be obtained from Eqs.~(\ref{eq:Notation Structure Type I}) and (\ref{eq:DW Energy Type I}):
\begin{eqnarray}
 \sigma_{\zeta}\left(\pi\over 4\right)= \eta^3_{0}\sqrt{\beta_{2}(K_{1}+K_{2})\over 2}\int^\pi_0d\gamma  \sin \gamma \sqrt{1\mp q \sin\gamma)}
\nonumber \\
= \eta^3_{0}\sqrt{2\beta_{2}(K_{1}+K_{2})}\left[
_3 F_2\left(-{1\over 4},{1\over 4},1;  {1\over 2},{3\over 2}; q^2\right)
 \mp
{\pi q\over 8}\left._2 F_1\left({1\over 4},{3\over 4};2;q^2\right)\right.\right],
\end{eqnarray}
where
\begin{equation}
  q=\frac{K_{3}+K_{4}}{K_{1}+K_{2}},
                                        \end{equation}
and $_pF_q(a_1,...,a_p; b_1,...,b_q;z)$ is the generalized hypergeometric function \cite{TF}.

\subsection{\label{RofOP}Rotation of order parameter: $\Theta$-wall}

Let us consider now the DWs with rotation of the order parameter
${\mathbf \eta}$ in the configuration space. Inside this domain wall
the phase $\Theta$ rotates from $-\pi/4$ to $+\pi/4$
(counterclockwise rotation) or  from $-\pi/4$ to $-3\pi/4$
(clockwise rotation).  We address this type of the DW as
$\Theta$-wall. The whole analysis, which was performed above for the
$\zeta$-wall, can be repeated for the $\Theta$-wall. This yields the
expressions obtained from those for the $\zeta$-wall by the
following substitution: $ \zeta \to 2\Theta$, $\psi \to \psi+\pi/4$,
$K_{1}-K_{2}\rightleftarrows K_{3}+K_{4}$ and $\beta_{2} \to
\beta_{2}-4\beta_{3}$. In particular, the final expressions for the
magnetic-induction jump and the DW surface energy are:

\begin{equation}
 \delta B_{\Theta}(\psi)=-(4\pi/c)\int dx' j_{y'}=\frac{8\pi e}{c}\eta^{2}_{0}\left\{ (K_{3}-K_{4}) \mp \frac{K_{3}+K_{4}}{\cos 2\psi} \left[\frac{\pi}{2}-\sin ^{2} 2\psi \frac{2}{(1-q^{2})^{1/2}}\arctan\sqrt{\frac{1\mp q}{1\pm q}}\right]\right\},
 \label{eq:Jump of magnetic induction Type II}
\end{equation}
where
\begin{equation}
  q=\frac{K_{1}-K_{2}}{K_{1}+K_{2}}\cos 2\psi,
                                        \end{equation}
and
\begin{equation}
 \sigma_{\Theta}(\psi)=\eta^{3}_{0}[2\beta_{2}(K_{1}+K_{2})]^{1/2}\int^\pi_0d\gamma  \sin \gamma \sqrt{f(\gamma)},
 \label{eq:DW Energy Type II}
\end{equation}
where
\begin{equation}
 f(\gamma)=1\mp \frac{K_{1}-K_{2}}{K_{1}+K_{2}}\cos 2\psi \sin \gamma-\left(\frac{K_{3}+K_{4}}{K_{1}+K_{2}}\right )^{2}{\sin^{2}2\psi \over 1\pm q \sin \gamma}.
 \label{eq:Notation Structure Type II}
\end{equation}
Here  $\gamma =2\Theta +\pi/2$ or $\gamma =-2\Theta -\pi/2$
for $\Theta$--rotations from $-\pi/4$ to $\pi/4$ or to $-3\pi/4$ respectively.   The DW surface energy has extrema at $\psi=0$ where the magnetic-induction jump is
\begin{equation}
\delta B_{\Theta}(0)=\frac{8\pi e}{c}\eta^{2}_{0}\left[K_3-K_4 \mp \frac{\pi}{2}(K_3+K_4)\right].
\end{equation}

\section{Discussion}

Let us discuss consequences of the obtained analytical and numerical
results. We start from the case of full axial symmetry in the $xy$
plane, when $K_{1}=K_{2}+K_{3}+K_{4}$ and $\beta_3=0$. In this case
the $\psi$ dependencies for the $\zeta$- and the $\Theta$-walls are
identical except for the shift of $\psi$ by $\pi/4$. Naturally the
choice of crystallographic axes does not matter in this case, only
their orientation with respect to the DW being important. The
minimum of the surface energy for the $\zeta$-wall corresponds (see
Fig.~\ref{Fig_1}b) to the angle $\psi=\pm \pi/4$ (45$^\circ$)
between the DW and the axes for $\mathbf \eta$. However, choosing
for  $\mathbf \eta$ the coordinate system connected with the DW, the
DW becomes  a $\Theta$-wall.

Let us consider now the tetragonal symmetry with $\beta_3 \neq 0$. Without any loss of generality one may assume that $\beta_3<0$. Indeed, if $\beta_3 >0$ with $\pi/4$-rotation of the crystal axes one obtain the 4th order terms with $\beta_2 -\beta_3/4$ instead of $\beta_2$ and $-\beta_3$ instead of $\beta_3$. Thus after this rotation the new $\beta_3$ is negative. At our choice of crystallographic axes the $\zeta$-wall has a smaller surface energy with the minimum at $\psi=\pm \pi /4$. Thus the stable DW is a $\zeta$-wall with respect to the order parameter axes, but is a $\Theta$-wall, if one use the order parameter ${\mathbf \eta}$ defined in the coordinates related to the stable DW.  The stable configuration of the DW corresponds to the maximum jump of the magnetic induction given by Eq. ~(\ref{eq:Maximum Jump}) with the lower sign before the second term.  On the other hand, the jump is minimal and is given by Eq.~(\ref{eq:SSE and Jump I}) if the DW is parallel to one of the two crystallographic axes.
It is interesting to compare these extremal  values  with the  first
critical magnetic field  given by Eq.~(\ref{Hc1}):
\begin{equation}
{\delta B_{max}\over H_{c1}} = \frac{\pi(K_{1}-K_{2}) }{ (K_1+K_2) \ln (\lambda/\xi) },~~
{\delta B_{min}\over H_{c1}} = \frac{4(K_{1}K_4-K_{2}K_3) }{ (K_1+K_2)^2 \ln (\lambda/\xi) }.
\end{equation}
For the values $\lambda=190$ nm and $\xi=66$ nm \cite{Kirtley} and assuming the weak-coupling relations $K_1/3=K_2=K_3=K_4$, these ratios are  1.48 and 0.48 respectively.

Since now there is no freedom for the choice of the axes for the order parameter ${\mathbf \eta}$ there is a force, which tends to orient the DW at 45$^\circ$ to the crystallographic axes. This  force may compete with other forces on the DW, e.g. , those from the sample shape. This would result in various configurations of DW not necessary those dictated by the bulk tetragonal anisotropy.

Whereas in the Landau-Lifshitz theory of ferromagnetism the
magnetic-induction jump is given by the universal value $8\pi M$,
where $M$ is the local spontaneous magnetization, in our case this
jump depends on the structure and the orientation of the DW.
Moreover, if one tried to introduce formally the effective
magnetization via the relation $\tilde M =\delta B/8\pi$ the latter
has no straightforward connection with the orbital moment $\mathbf
l$ of the Cooper pair and even can have a negative sign with respect
to the intrinsic magnetic moment of the Cooper pair.

Returning back to the question why they could not detect stray fields generated by the magnetic-induction jumps outside  $\textrm{Sr}_{2}\textrm{RuO}_{4}$ samples \cite{Kirtley} our analysis demonstrates that at the present stage the theory cannot make definite predictions on the strength of these fields without reliable information on the parameters in the GL theory.
In principle, one could suggest that the DWs in the experiments were not in the ground state, or the GL parameters are essentially different from their weak-coupling-limit values. But further experimental and theoretical work is needed to check these suggestions.

\section{\label{Conclusions}Conclusions}

We investigated properties of DWs in a tetragonal chiral $p$-wave
superconductor. Various cases of the DW structure and orientation
with respect to in-plane crystallographic axes were analyzed. The
magnetic-induction jump on the DW changes from case to case, in
contrast to the Landau-Lifshitz theory of ferromagnetism, where this
jump has the universal value proportional to the local spontaneous
magnetization. This conclusion stresses again that the latter is not
defined for orbital ferromagnetism. The  magnetic-induction jump is
responsible for stray magnetic fields, which have not yet detected
outside $\textrm{Sr}_{2}\textrm{RuO}_{4}$ samples \cite{Kirtley}. A
quantitative evaluation of stray fields cannot be done without a
more detailed information (from the microscopical theory or from the
experiment) of the parameters of the GL theory.

\section*{Acknowledgments}

This work has been supported by the grant of the Israel Academy of
Sciences and Humanities.

  \end{document}